\documentclass[%
 reprint,
 amsmath,amssymb,
 aps,
]{revtex4-2}

\usepackage{graphicx}
\usepackage{dcolumn}
\usepackage{bm}
\usepackage{amssymb,amsmath,amsthm,mathrsfs,color,MnSymbol,stmaryrd,enumitem,hyperref,float, accents,comment,subcaption,graphicx,physics,tikz}
\graphicspath{ {./figures/} }

\newcommand{\myequation}{\begin{equation}}
\newcommand{\myendequation}{\end{equation}}
\let\[\myequation
\let\]\myendequation

\begin{document}

\preprint{APS/123-QED}

\title{A Novel Method of Function Extrapolation Inspired by Techniques in Low-entangled Many-body Physics}


\author{Lambert Lin}
\author{Steven R White}
\email{srwhite@uci.edu}
\affiliation{Department of Physics, University of California, Irvine, CA 92697, USA}

\date{\today}

\begin{abstract}
We introduce a novel extrapolation algorithm inspired by quantum mechanics and evaluate its performance against linear prediction. Our method involves mapping function values onto a quantum state and estimating future function values by minimizing entanglement entropy. We demonstrate the effectiveness of our approach on various simple functions, both with and without noise, comparing it to linear prediction. Our results show that the proposed algorithm produces extrapolations comparable to linear prediction, while exhibiting improved performance for functions with sharp features.
\end{abstract}

\maketitle


\section{Introduction}

Extrapolation algorithms are recipes to estimate future values given past data as input, or values outside an interval based on values in an interval \cite{Extrapolationmethods,numericalrecipes}. Typically the algorithms can be iterated to provide a sequence of values indefinitely. Given a dataset, observed values $y_1, y_2, ..., y_n$, which represents the values of a function $f(x)$ evaluated at $x_1, x_2, ..., x_n$ with $x_1 < x_2 < ... < x_n$, the goal is to predict $f(x)$ for arbitrary values of $x$ outside the interval $[x_1, x_n]$ based on observed trends and patterns \cite{Extrapolationmethods,numericalrecipes}.
A primary objective is to maintain the stability of the estimates during iteration \cite{practical} while maintaining accuracy \cite{accuracy}. Various successful extrapolation algorithms have been developed, including polynomial regression, spline extrapolation, and linear prediction \cite{Extrapolationmethods,numericalrecipes,vectorsequences}. However, each algorithm has inherent limitations, excelling in certain types of datasets but failing to produce reliable predictions universally \cite{AdvantagesandLimitations}. Therefore, it is crucial to identify which existing algorithms work well for specific datasets that require extrapolation.

Extrapolation finds applications in diverse domains, including quantitative analysis \cite{finacial}, weather forecasting \cite{weather}, demographics, and more \cite{spectral,wildlife}. It is commonly employed in predicting future revenue and stock prices based on historical data \cite{stock}, as well as anticipating weather changes and population growth. By extending past patterns into the future, extrapolation helps to understand and anticipate future behavior when relying solely on historical patterns is insufficient.

In this paper, we introduce a novel extrapolation algorithm called entropy extrapolation. Inspired by quantum mechanics, this algorithm leverages the encoding of classical functions into quantum states \cite{tensortrain} using an orthonormal basis of qubits. By utilizing quantum operations, entropy extrapolation enhances the estimation of function values, sidestepping classical limitations. The encoded state is entangled, and the entanglement entropy serves as a measure of the degree of entanglement \cite{arealaw,quantumentangle}. Our algorithm predicts a function value at a future point by minimizing the local entanglement entropy. Entropy extrapolation demonstrates particular effectiveness for smooth functions, as the resulting quantum states tend to exhibit low entanglement. To provide a comparative analysis, we will compare entropy extrapolation with the well-established linear prediction method. For simplicity, our focus will be on the case when the $x_i$ values are equally spaced.

In the upcoming section, we provide an overview of the linear prediction method, which will serve as our primary benchmark. In Section III, we introduce the entropy extrapolation method and present its mathematical and computational formulations. Subsequently, in Section IV, we apply the entropy extrapolation technique to a range of functions, followed by a comparative analysis of the results alongside those obtained using linear prediction.

\section{Linear Prediction}
\subsection{Introduction} 
Linear prediction is an extrapolation algorithm commonly used to estimate future function values. It predicts $y_{n+1}$, the function value $f(x)$ at $x_{n+1}$, as a linear combination of the previous $m$ values \cite{Linearprediction}, given by:
\[y_{n+1} = \sum_{j=1}^{m} y_{n+1-j}d_j \:.\]
Here, $m$ represents the order of linear prediction coefficients, which determines the number of previous values used to estimate the next value. The linear prediction coefficients, denoted as $d_j$, are computed based on the available data \cite{Linearprediction,numericalrecipes}.

\subsection{Linear prediction coefficients}
The linear prediction coefficients are obtained by minimizing the sum of squared errors. The error for a point $y_i$ is defined as the difference between the exact value and the estimated value $y_i - \sum_{j=1}^{m}y_{i-j}d_j$. The mathematical derivation is based on the approach presented in the renowned "Numerical Recipes" by Press et al.\:\cite{numericalrecipes}. However, the treatment in the book is schematic and does not correctly describe a finite range of initial data.

In the case of linear prediction as an extrapolation algorithm, we estimate the subsequent value based on the previous $m$ values. The first value for which the algorithm provides an estimation is $f(x_{m+1})$. Consequently, the errors for $y_1,...,y_m$ cannot be defined and must be excluded from the squared errors.

To generalize the approach to handle complex-valued data sets, we modified the derivation. For a complex array $y_1,...,y_n$ with $y_i \in \mathbb{C}$, the squared error is given by:
\begin{align}
    S =& \sum_{i=m+1}^{n} \lvert y_i - \sum_{j=1}^m y_{i-j}d_j \rvert^2\\\nonumber
    =&\sum_{i=m+1}^{n} \lvert y_i \rvert^2 - \sum_{i=m+1}^{n} \sum_{j=1}^m (y_i^* y_{i-j} d_j -  y_i y_{i-j}^* d_j^*)  \\\nonumber & \ + \sum_{i=m+1}^{n} \sum_{j=1}^m \sum_{k=1}^m y_{i-j}y_{i-k}^* d_j d_k^* \:.
\end{align}
To minimize the sum of squared errors, we set the partial derivative of $S$ with respect to $d_j $ to 0:
\begin{equation}
\frac{\partial S}{\partial d_j} = -\sum_{i=m+1}^{n} y_i^* y_{i-j} + \sum_{i=m+1}^{n} \sum_{k=1}^m  y_{i-j}y_{i-k}^*d_k^* = 0 \:.
\end{equation}
We can express Eq.~(3) with matrices and vectors, setting $\phi_j = \sum_{i=m+1}^{n} y_i^* y_{i-j}$ and $M_{jk} = \sum_{i=m+1}^{n} y_{i-j}y_{i-k}^*$ and it becomes
\begin{equation}
    -\phi_j + M_{jk}d_k^* = 0\:.
\end{equation}
The linear prediction coefficients are solved as
\begin{equation}
    d_k = (M_{jk}^{-1}\phi_j)^*\:.
\end{equation}
Once we obtain $d_j$, we can estimate subsequent function values with Eq.~(1).

\subsection{Discussion on performance}
Linear prediction exhibits excellent performance on periodic analytic functions, particularly plane waves and sinusoidal functions \cite{periodic}. For instance, when considering the plane wave function $f(x) = e^{ikx}$, it can be analytically justified that linear prediction with a single coefficient (order 1) yields perfect estimation. In this case, the coefficient $d$ represents a constant phase $e^{ik\Delta x}$, where $\Delta x$ is the spacing between $x_i$ values. Multiplying $d$ with $f(x)$ produces the same result as evaluating $f(x+\Delta x)$.
\begin{figure}[!htb]
     \centering
     \begin{subfigure}[b]{0.43\textwidth}
         \centering
         \begin{tikzpicture}
         \draw (0, 0) node[inner sep=0] {\includegraphics[width=\textwidth,trim={15mm 0 15mm 0},clip]{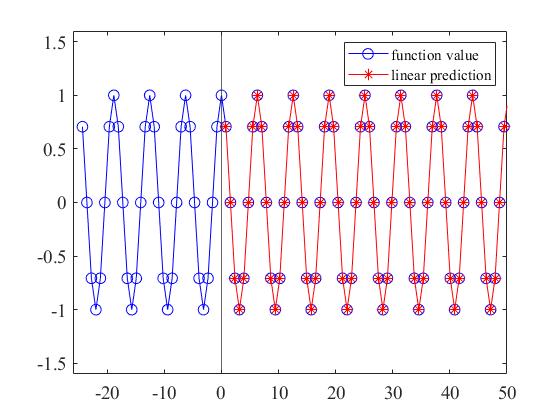}};
         \draw (-2.5, 2.3) node {(a)};
         \end{tikzpicture}
     \end{subfigure}
     \begin{subfigure}[b]{0.43\textwidth}
         \centering
         \begin{tikzpicture}
         \draw (0, 0) node[inner sep=0] {\includegraphics[width=\textwidth,trim={15mm 0 15mm 0},clip]{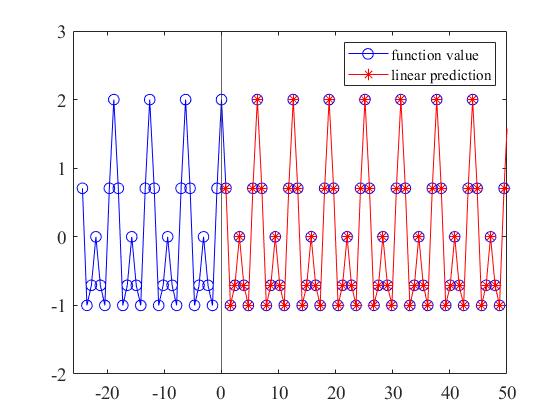}};
         \draw (-2.5, 2.3) node {(b)};
         \end{tikzpicture}
     \end{subfigure}
     \begin{subfigure}[b]{0.43\textwidth}
         \centering
         \begin{tikzpicture}
         \draw (0, 0) node[inner sep=0] {\includegraphics[width=\textwidth,trim={15mm 0 15mm 0},clip]{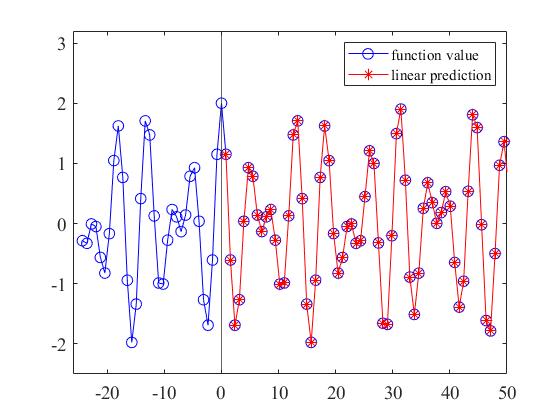}};
         \draw (-2.5, 2.3) node {(c)};
         \end{tikzpicture}
     \end{subfigure}
     \caption{\raggedright Estimation results of plane wave functions with linear prediction. (a). $f(x)=e^{ix}$ estimated with $m=1$. (b). $f(x)=e^{ix}+e^{2ix}$ estimated with $m=2$. (c). $f(x) = \cos x +\cos \sqrt{2} x$ estimated with $m=4$.}
        \label{fig:1}
\end{figure}
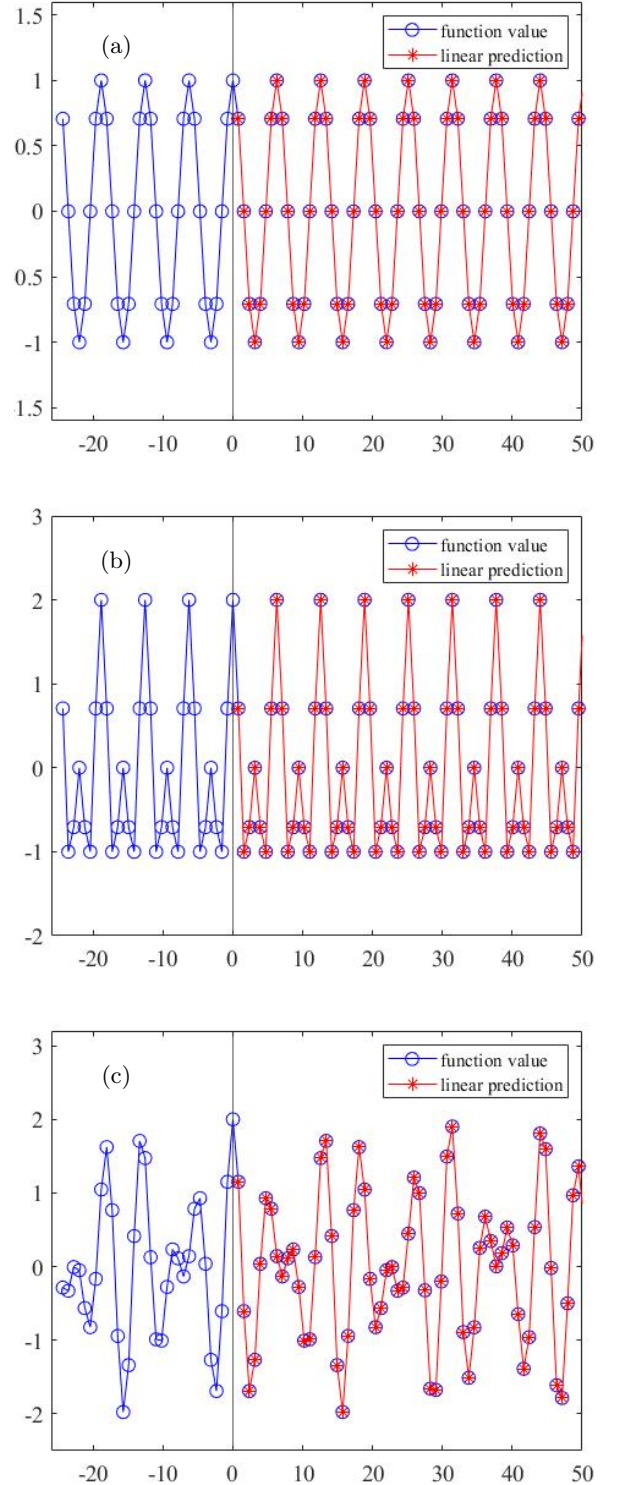


This conclusion can be extended to sums of plane waves with different frequencies. Each additional frequency necessitates an extra prediction coefficient to achieve perfect estimation. Fig.~\ref{fig:1} illustrates linear prediction results for plane wave functions with varying frequencies, serving as examples. The extrapolation starts at the black vertical line, with the input function values for the algorithm shown in blue, and the output estimation results depicted in red. The figures only display the real part of the complex function values.

\begin{figure}[!htb] 
     \centering
     \begin{subfigure}[b]{0.43\textwidth}
         \centering
         \begin{tikzpicture}
         \draw (0, 0) node[inner sep=0] {\includegraphics[width=\textwidth,trim={15mm 0 15mm 0},clip]{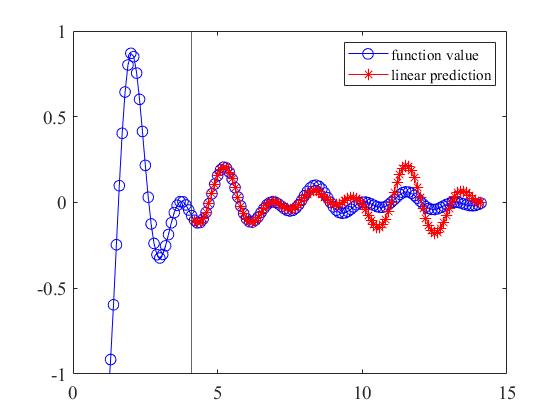}};
         \draw (-2.9, 2) node {(A)};
         \end{tikzpicture}
         \caption{\label{first}}
     \end{subfigure}
     \begin{subfigure}[b]{0.43\textwidth}
         \centering
         \begin{tikzpicture}
         \draw (0, 0) node[inner sep=0] {\includegraphics[width=\textwidth,trim={15mm 0 15mm 0},clip]{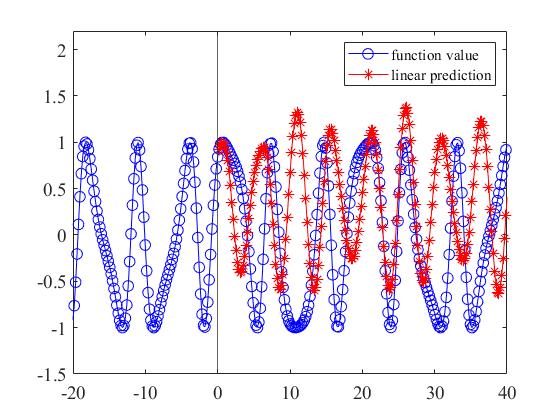}};
         \draw (-2.5, 2) node {(B)};
         \end{tikzpicture}
         \label{fig:y equals x}
     \end{subfigure}
        \caption{\raggedright Linear prediction results of bounded non-periodic functions. (A) $f(x) = \mathcal{F}^{-1}[\tilde{f}(\omega)]$, where $\tilde{f}(\omega) = \sqrt{1-(\omega-3)^2}$ for $2\le \omega \le 4$ and zero otherwise. (B) $f(x) = \sin(x+\cos(\frac{x}{\sqrt{2}}))$.}
        \label{fig:2}
\end{figure}

However, linear prediction's performance diminishes for non-analytic periodic functions \cite{sloppiness} such as square waves and sawtooth functions. These functions can be expressed as Fourier series, comprising an infinite sum of plane waves with different frequencies. When extrapolating with a small value of $m$, linear prediction provides an approximate estimation as a finite sum of leading sinusoidal terms. Nevertheless, extrapolating with a large $m$ includes higher-order terms, which can be affected by the noise present in the provided samples \cite{noise,sloppiness}, leading to significant deviations in the estimation. Similarly, linear prediction does not fare well with bounded non-periodic functions. Fig.~\ref{fig:2} demonstrates two examples of linear prediction applied to functions of this type. The first example is inspired by a problem in many-body physics, where the function in the frequency domain represents a half unit circle centered at $\omega = 3$. Upon performing the inverse Fourier transform, it transforms into a bounded and non-periodic function in the position domain, providing a relevant illustration for our discussion.

\subsection{Stability}
In certain scenarios, linear prediction as an extrapolation algorithm may result in estimated values with rapidly increasing amplitudes that tend towards infinity. This instability arises due to the fact that linear prediction is not universally stable for all types of functions. However, we can address this issue by applying an additional step to ensure stability \cite{numericalrecipes}. The stability condition is determined by examining the characteristic polynomial of the linear prediction coefficients:
 \[ z^N - \sum_{j=1}^N d_jz^{N-j} = 0\:.\]
For stability, all roots of the polynomial must reside within the unit circle, i.e., $|z| \le 1$. In cases where instability is observed, we can follow the following standard procedure to rectify it:

1) Normalize the unstable roots by mapping them onto the unit circle.

2) Recalculate the modified linear prediction coefficients, denoted as $d_j\,'$, using the adjusted roots.

Once we have the modified coefficients $d_j\,'$, we can perform a stable estimation of our function using these new coefficients.

 \section{entropy extrapolation}
\subsection{Motivation}
While linear prediction excels in estimating sinusoidal functions, it demonstrates clear limitations when applied to non-analytic functions. Classical extrapolation algorithms face similar challenges \cite{classical} in this context, prompting the search for alternative approaches. In this regard, the principles of quantum mechanics offer a promising direction for problem-solving. By leveraging the conversion of classical function values into quantum states, we can employ quantum techniques to tackle the problem, overcoming the limitations of classical methods. Tensor networks and matrix product states, which are effective tools \cite{classical} for analyzing quantum systems with low entanglement \cite{tensornetwork}, have been successfully applied to study encoded classical periodic functions. This suggests the possibility of developing an extrapolation algorithm that leverages quantum techniques to achieve outstanding performance on such functions.
\subsection{Encoding}
We aim to represent the vector $y_i = (y_1, y_2, ..., y_n)$ using an entangled quantum state within the orthonormal basis $\{\ket{0}, \ket{1}\}$. This representation of the vector $y$ is commonly referred to as the Quantized Tensor Train (QTT) representation. For instance, a three-qubit state can be expressed as $\ket{\psi} = \sum_{i,j,k = 0,1} c_{ijk}\ket{i}\ket{j}\ket{k}$, where the coefficients $c_{ijk}$ correspond to a (0, 3)-tensor comprising complex entries. To achieve the representation of the vector $y_i$ through the quantum state $\ket{\psi}$, it is necessary to establish a mapping from $y$ to the tensor $c$, which is denoted as the tensor train format of the vector $y_i$.

The tensor train format is constructed by reshaping a vector with $2^n$ entries into a (0, n)-tensor. For instance, if $y_i = (y_1,y_2,...,y_8)$, we reshape it into a (0, 3)-tensor $c_{ijk}$ with the assignment:
\[c_{0jk}=
\begin{bmatrix}
y_1 & y_2 \\
y_3 & y_4 
\end{bmatrix},\ 
c_{1jk}=\begin{bmatrix}
y_5 & y_6 \\
y_7 & y_8 
\end{bmatrix}.\]

An alternative way to comprehend the reshaping process is to convert $i-1$ into a binary number, which serves as the tensor index, and map the $i$th entry of the vector to the corresponding tensor entry. For instance, $c_{100} = y_5$ since 4 (derived from $5-1$) is expressed as 100 in binary notation. Reshaping can be easily accomplished in computer programs using the built-in reshape function.

\subsection{Matrix product state}
A matrix product state (MPS) represents the factorization of a (0, n)-tensor into a product of matrices \cite{tensornetworkmethods}, providing an efficient description of quantum systems with low entanglement. This technique can be leveraged to process data that is mapped to a low-entangled state. A tensor $T_{a_1a_2,...,a_n}$ can be written in MPS form as:
\[T_{a_1a_2,...,a_n} = A_1A_2...A_n = \sum_{\alpha}A_1^{\alpha_1}A_2^{\alpha_1\alpha_2}...A_n^{\alpha_{n-1}},\]
where $A_1, A_2, ..., A_n$ are complex matrices of rank $d$, and $\alpha$ denotes the bond indices summed over. In the tensor network language, $d$ is referred to as the bond dimension, and it can vary from bond to bond \cite{mps}.
\begin{figure}[!htb]
     \centering
     \includegraphics[width=0.45\textwidth]{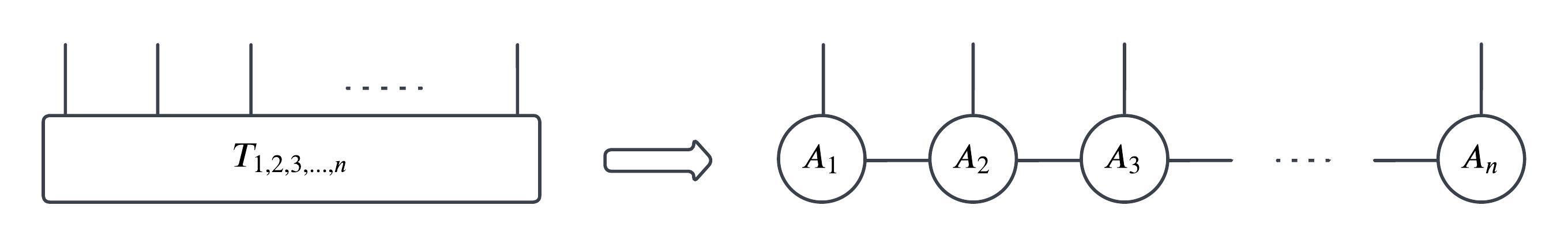}
     \caption{\raggedright The process of constructing the MPS from a tensor expressed in tensor diagram notation.}
     \label{fig:MPS}
\end{figure}
The bond index $\alpha$ represents the dimensionality of the connection between successive matrices in the MPS chain \cite{dmrg}. The MPS can also be represented using tensor diagram notation \cite{geometry}, as shown in Fig.~\ref{fig:MPS}. The vertical lines correspond to the uncontracted indices $a_1, a_2, ..., a_n$, while the horizontal lines depict the contracted bonds.

Singular value decomposition (SVD) is commonly used to express a tensor as an MPS \cite{svd}. The tensor $T_{a_1a_2,...,a_n}$ can be decomposed as:
\begin{align}
    T_{a_1a_2,...,a_n} &= A_1S_{12}T'_{a_2,...,a_n} \\\nonumber
    &= A_1S_{12}A_2S_{23}T''_{a_3,...,a_n} \\\nonumber
    &= ...\\\nonumber
    &= A_1S_{12}A_2S_{23}...S_{n-1,n}A_n\:,
\end{align}
where the $S$ matrices are singular value matrices \cite{svdapp}. The complete decomposition of a tensor into an MPS involves repeated application of SVD. 
\begin{figure}[!htb]
     \centering
     \includegraphics[width=0.45\textwidth]{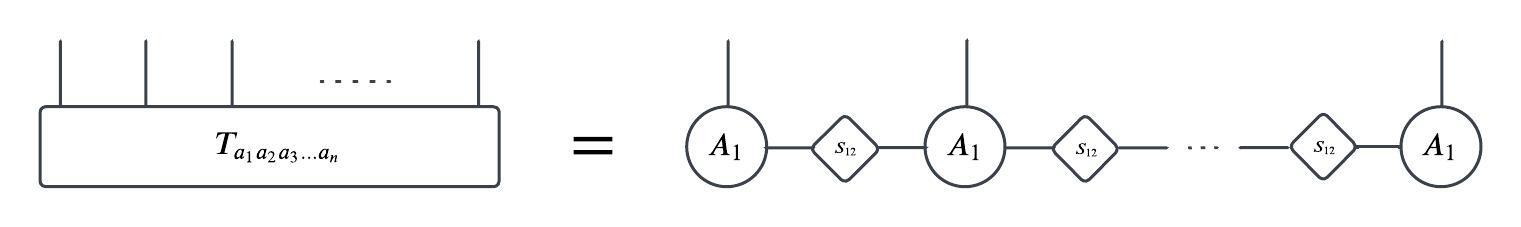}
     \caption{\raggedright Constructing MPS with SVD decomposition. $S_{ij}$ is the singular value matrix on the bond that connects $A_i$ to $A_j$.}
     \label{fig:SVD}
\end{figure}
This process is depicted in the tensor diagram shown in Fig.~\ref{fig:SVD}. The singular value matrices can be absorbed into either the left or right $A$ matrices to obtain the form shown in Fig.~\ref{fig:MPS}. However, for the purpose of our discussion, we maintain the explicit representation of $S$ matrices.

\subsection{Entanglement and entanglement entropy}
The degree of entanglement in a quantum state can be characterized by the bond dimension of its MPS \cite{area,arealaw}. A state with higher bond dimensions exhibits more significant entanglement, which corresponds to larger entanglement entropy. The entanglement entropy of a bond cut can be directly calculated from the corresponding singular value matrix, and the total entanglement entropy of the state is obtained by summing the entropies of all bonds. Suppose we have a singular value matrix $S$ with $n$ singular values, which can be written as a vector $s_i = (s_1, s_2, ..., s_m)$. By dividing each $s_i$ by its norm $\lvert s \rvert$, we obtain the normalized singular values, which represent the square roots of the probabilities:
\[p_i = \frac{s_i^2}{\lvert s \rvert^2}\:.\]
It is important to note that the probabilities sum up to 1. Using these probabilities, the entropy of the bond cut can be calculated according to the definition. In our algorithm, we employ the more generalized definition proposed by R\'enyi, as opposed to the commonly used Shannon entropy \cite{renyi}. The R\'enyi entropy is defined as:
\[H(\alpha) = \frac{1}{1-\alpha} \ln (\sum_{i=1}^m p_i^{\alpha})\:,\]
where $\alpha \ge 0$ and $\alpha \neq 1$. The limit as $\alpha \to 1$ corresponds to the Shannon entropy, while the limit as $\alpha \to \infty$ corresponds to the min-entropy $H_{\infty}$, which is determined by the highest possible event. By evaluating the entropy of each bond using Eq.~(11) and summing them up, we obtain the total R\'enyi entropy of an MPS.

Our approach to estimating future values of classical functions capitalizes on the relationship between entropy and the entanglement of states to which these functions are mapped. We are particularly interested in functions that can be mapped to low-entangled states. Examples of such functions include simple periodic functions like sinusoidal functions, sawtooth waves, square waves, and others. Certain exponential functions can also be mapped perfectly to a state with a bond dimension of 2 \cite{tensortrain}. It is reasonable to assume that functions exhibiting low entanglement in their mapped states will continue to maintain this property. Thus, we are likely to find potential future values within regions where the state remains low-entangled. The relation between entropy and entanglement suggests that low-entangled regions correspond to minimum entropy. However, further investigation is needed to understand which specific functions can be mapped to low-entangled states and establish a detailed classification standard.

From this perspective, we introduce an extrapolation algorithm called entropy extrapolation. Given a sequence of function values $y_1, y_2, ..., y_n$ and the desire to estimate $y_{n+1}$, we map the last $2^q$ values into a $q$-qubit state, where the entropy of the state becomes a function of $y_{n+1}$. We estimate $y_{n+1}$ to be the value that minimizes the entropy of the state. This method can be iteratively applied to predict subsequent function values.

The parameter $\alpha$ influences the precision of entropy extrapolation. Setting $\alpha$ to be either too large or too small can increase the algorithm's instability. Based on our testing examples, we find that $\alpha$ values ranging from 0.2 to 0.3 generally yield stable and accurate estimations in most cases. However, the range of $\alpha$ for generating precise estimations may vary for different function types. Therefore, it is advisable to experiment with different $\alpha$ values when applying this algorithm.

\subsection{Entropy behavior at minimum}
To gain a deeper understanding of the behavior of entropy extrapolation, it is crucial to explore how the entropy behaves in the immediate vicinity of its minimum. Such an exploration provides valuable insights into the operational dynamics of the algorithm and the underlying reasons for its convergent behavior.

The appendix offers an in-depth elucidation of the entropy behavior around its minimum, along with concrete examples that provide tangible insights. These instances vividly illustrate that the entropy swiftly approaches its minimum and adopts a sharply pointed configuration at this nadir as shown in Fig.~\ref{app2}. This phenomenon substantiates the efficacy of entropy extrapolation in accurately estimating function values, as the actual function value is anticipated to closely align with the estimated value at the minimum. This proximity ensures the maintenance of a low-entangled state, validating the algorithm's robustness and accuracy.

\begin{figure}[h]
     \centering
     \begin{subfigure}[b]{0.43\textwidth}
         \centering
         \begin{tikzpicture}
         \draw (0, 0) node[inner sep=0] {\includegraphics[width=\textwidth,trim={15mm 0 15mm 0},clip]{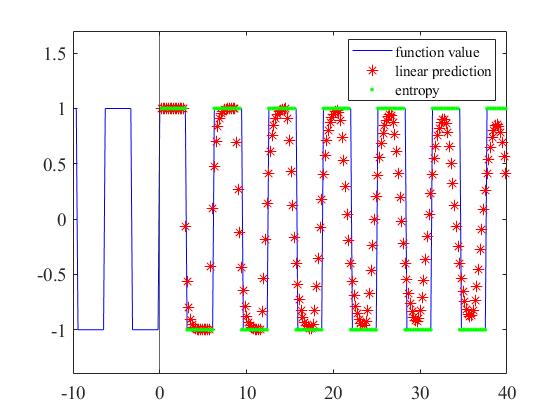}};
         \draw (-2.5, 2) node {(A)};
         \end{tikzpicture}
         \label{fig:y equals x}
     \end{subfigure}
     \hspace{0.5cm}
     \begin{subfigure}[b]{0.43\textwidth}
         \centering
         \begin{tikzpicture}
         \draw (0, 0) node[inner sep=0] {\includegraphics[width=\textwidth,trim={15mm 0 15mm 0},clip]{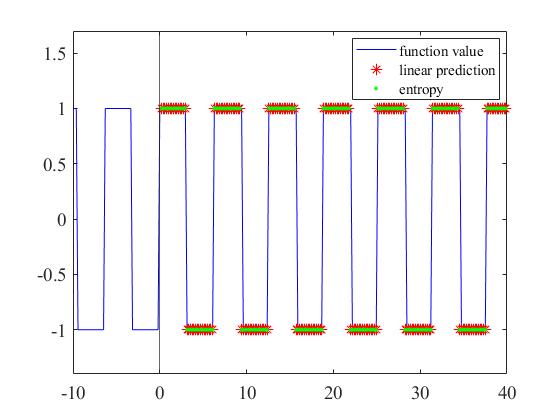}};
         \draw (-2.5, 2) node {(B)};
         \end{tikzpicture}
         \label{fig:y equals x}
     \end{subfigure}
        \caption{\raggedright Square wave extrapolated with linear prediction and entropy extrapolation (A) $m = 15$ for linear prediction and $q = 7$ for entropy extrapolation (B) $m = 30$ for linear prediction and $q = 7$ for entropy extrapolation.}
        \label{fig:6}
\end{figure}

\section{examples and comparison with linear prediction} 
We conducted several experiments to evaluate the accuracy of entropy extrapolation in estimating different functions. In this section, we present examples of entropy extrapolation and compare its performance to linear prediction. Additionally, we apply both methods to cases where noise is present in the provided data and analyze the performance of entropy extrapolation in noisy scenarios. In the examples provided below, we use $\alpha = 0.25$ to calculate the R\'enyi entropy.

Fig.~\ref{fig:6} displays the extrapolation results of the square wave for both entropy 
\begin{figure}[htb!]
     \centering
     \includegraphics[width=0.43\textwidth,trim={15mm 0 15mm 0},clip]{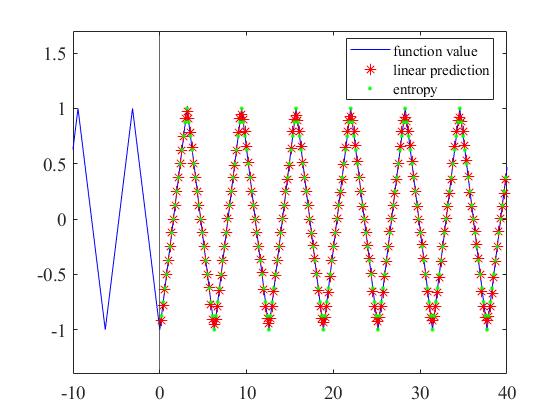}
     \caption{\raggedright Sawtooth function extrapolated with linear prediction ($m=15$) and entropy extrapolation ($q=7$).}
     \label{fig:7}
\end{figure}

\begin{figure}[htb!]
     \centering
     \includegraphics[width=0.43\textwidth,trim={15mm 0 15mm 0},clip]{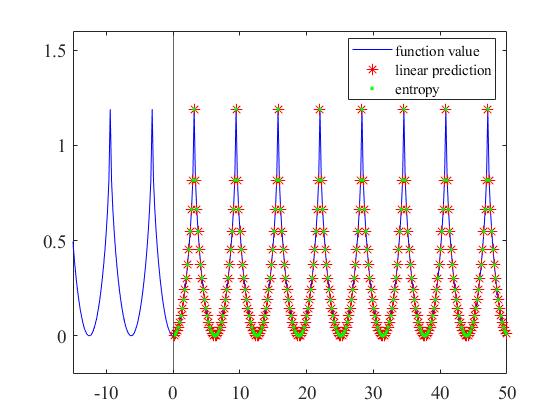}
     \caption{\raggedright $f(x) = 2^{\frac{1}{4}}- (1+ \cos x)^{\frac{1}{4}}$ extrapolated with linear prediction ($m=50$) and entropy extrapolation ($q=7$).}
     \label{fig:10}
\end{figure}
extrapolation and linear prediction in the absence of noise. Entropy extrapolation exhibits desirable estimation accuracy, while linear prediction only achieves good estimations of the square wave when the order is sufficiently high.

Two additional examples of periodic functions extrapolated using both methods are presented in Fig.~\ref{fig:7} and \ref{fig:10}. Linear prediction yields precise estimations only with very high orders in the case depicted in Fig.~\ref{fig:10}.

\begin{figure}[htb!]
     \centering
     \includegraphics[width=0.43\textwidth,trim={15mm 0 15mm 0},clip]{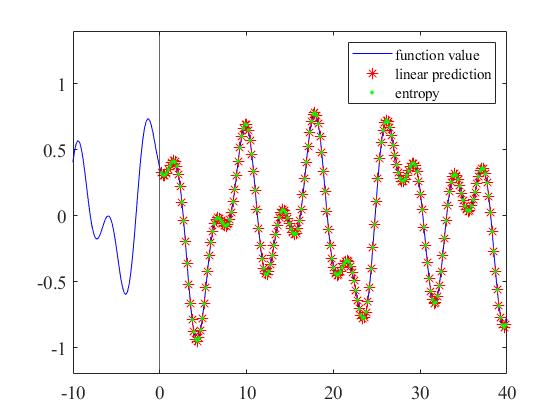}
     \caption{\raggedright Sum of multiple plane waves with random frequencies, amplitudes, and phases extrapolated with the linear prediction ($m=5$) and entropy extrapolation ($q=7$).}
     \label{fig:8}
\end{figure}

\begin{figure}[htb!]
     \centering
     \includegraphics[width=0.43\textwidth,trim={15mm 0 15mm 0},clip]{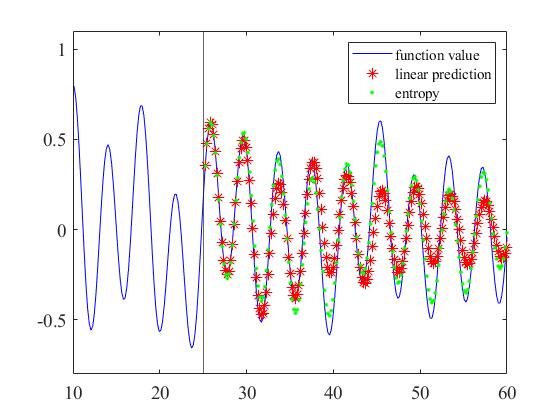}
     \caption{\raggedright Sum of plane waves with exponentially decaying amplitudes and random frequencies and phases extrapolated with the linear prediction ($m=5$) and entropy extrapolation ($q=7$).}
     \label{fig:9}
\end{figure}

Fig.~\ref{fig:8} displays the extrapolation results for the sum of multiple plane waves with random frequencies, while Fig.~\ref{fig:9} presents the extrapolation results for the sum of decaying plane waves with random frequencies. Neither method provides a highly accurate estimation of the decaying plane wave. As the number of plane waves increases, both methods generate less precise estimations of future values. To enhance precision, linear prediction necessitates a higher order, while entropy extrapolation requires more qubits. While it is well-understood that each plane wave requires 2 orders to achieve a good estimation with linear prediction, further investigation is needed to understand this pattern for entropy extrapolation. The mapping of the function into a low-entangled state significantly impacts the performance of entropy extrapolation. In the examples provided, we employ 7 qubits for entropy extrapolation, enabling accurate estimations of up to 4 plane waves with different frequencies.

\begin{figure}[h]
     \centering
     \begin{subfigure}[b]{0.43\textwidth}
         \centering
         \begin{tikzpicture}
         \draw (0, 0) node[inner sep=0] {\includegraphics[width=\textwidth,trim={15mm 0 15mm 0},clip]{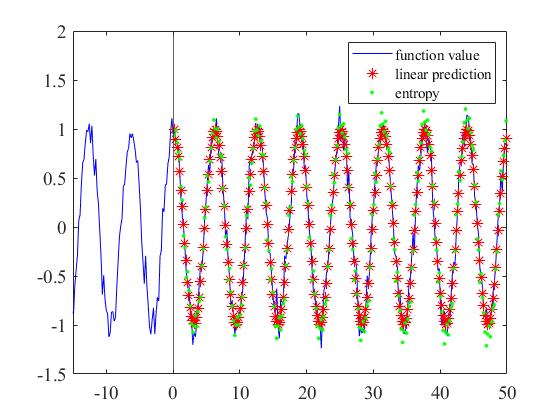}};
         \draw (-2.5, 2) node {(A)};
         \end{tikzpicture}
         \label{fig:y equals x}
     \end{subfigure}
     \hspace{0.5cm}
     \begin{subfigure}[b]{0.43\textwidth}
         \centering
         \begin{tikzpicture}
         \draw (0, 0) node[inner sep=0] {\includegraphics[width=\textwidth,trim={15mm 0 15mm 0},clip]{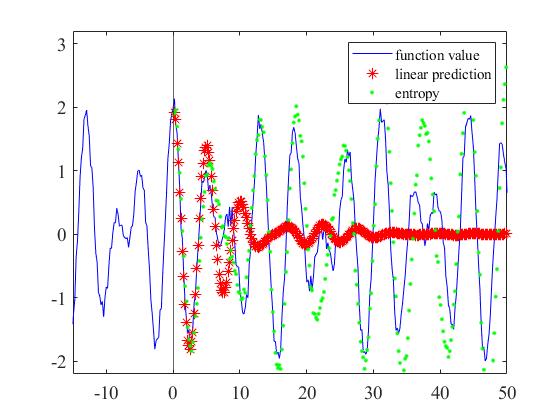}};
         \draw (-2.5, 2) node {(B)};
         \end{tikzpicture}
         \label{fig:y equals x}
     \end{subfigure}
        \caption{\raggedright Noisy sinusoidal functions extrapolated with linear prediction ($m=15$) and entropy extrapolation ($q=7$). The noise is added to the function (A) $f(x) = \cos(x)$ (B) $f(x) = \cos x+\cos \sqrt{2}x$.}
        \label{fig:11}
\end{figure}

Another crucial factor influencing the performance of extrapolation algorithms is the presence of noise in the data. Noise disrupts the pattern of the original data, thereby adversely affecting the accuracy of extrapolation. It is essential to assess how well entropy extrapolation can estimate noisy data, as real-world data often contains some level of noise. To examine the influence of noise on estimation results in specific cases, we introduce normally distributed noise to both periodic functions and sinusoidal functions, which are well-suited for both methods. Fig.~\ref{fig:11} illustrates the extrapolation results for noisy sinusoidal functions using linear prediction and entropy extrapolation. It is evident that linear prediction generates inaccurate estimations when the provided data is noisy. Conversely, noise has a relatively minor impact on entropy extrapolation.

\begin{figure}[h]
     \centering
     \begin{subfigure}[b]{0.43\textwidth}
         \centering
         \begin{tikzpicture}
         \draw (0, 0) node[inner sep=0] {\includegraphics[width=\textwidth,trim={15mm 0 15mm 0},clip]{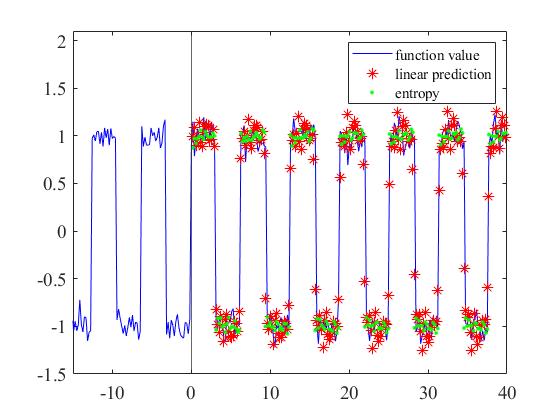}};
         \draw (-2.5, 2) node {(A)};
         \end{tikzpicture}
         \label{fig:y equals x}
     \end{subfigure}
     \hspace{0.5cm}
     \begin{subfigure}[b]{0.43\textwidth}
         \centering
         \begin{tikzpicture}
         \draw (0, 0) node[inner sep=0] {\includegraphics[width=\textwidth,trim={15mm 0 15mm 0},clip]{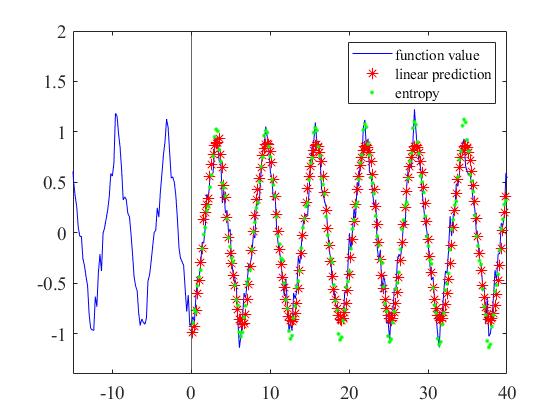}};
         \draw (-2.5, 2) node {(B)};
         \end{tikzpicture}
         \label{fig:y equals x}
     \end{subfigure}
        \caption{\raggedright Noisy periodic functions extrapolated with linear prediction ($m=25$) and entropy extrapolation ($q=7$). (A) Square wave (B) Sawtooth.}
        \label{fig:12}
\end{figure}

Similar observations can be made regarding linear prediction applied to noisy square waves, as depicted in Fig.~\ref{fig:12}. Linear prediction achieves higher accuracy in estimating noiseless data by utilizing a larger order $m$, which incorporates more data points for estimating subsequent values. However, in the presence of noise, the accuracy of linear prediction diminishes as noisy data points are included in the estimation process. The noise accumulates as the order $m$ increases, leading to decreased precision. In contrast, entropy extrapolation overcomes this limitation by being less affected by noise. The influence of noise on the entropy is relatively small, compared to its impact on the underlying pattern of the data. Thus, entropy extrapolation remains stable as long as the noisy data can be mapped into a low-entropy state.

\section{Conclusions}
In this study, we introduced a novel extrapolation method called entropy extrapolation, which draws inspiration from quantum many-body physics and tensor networks. This method applies to classical functions that can be mapped into low-entangled states. We evaluated the performance of entropy extrapolation on various periodic and non-periodic functions, including plane waves and square waves, with and without noise. Our findings indicate that entropy extrapolation demonstrates remarkable accuracy in estimating repeated patterns, even in the presence of noise. When compared to linear prediction, entropy extrapolation yields more precise and stable estimations for periodic functions affected by noise. However, further research is necessary to comprehensively understand the specific types of functions that can be effectively extrapolated using entropy extrapolation and explore methods to enhance its stability, such as modifications similar to those employed in linear prediction.

\begin{acknowledgments}
We thank Zhaoyi Li, Chris Chen, Peize Yu, and Giovanni Michel for their helpful discussions and insights. This work was supported in part by the National Science Foundation under grant DMR-2110041.
\end{acknowledgments}

\appendix*
\section{Entropy Behaviors}
\label{appdx}
This appendix offers additional insights into the entropy behavior within the search range of $y_{n+1}$, particularly in the proximity of the minimum. In the majority of scenarios, the entropy demonstrates a rapid convergence towards the minimum, often forming a distinctive sharp tip. 
\begin{figure}[htb!]
     \centering
     \begin{subfigure}[b]{0.43\textwidth}
         \centering
         \begin{tikzpicture}
         \draw (0, 0) node[inner sep=0] {\includegraphics[width=\textwidth,trim={15mm 0 15mm 0},clip]{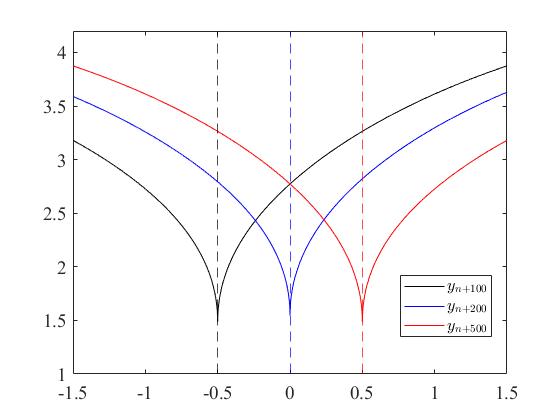}};
         \draw (-2.5, -2) node {(A)};
         \end{tikzpicture}
         \label{fig:y equals x}
     \end{subfigure}
     \hspace{0.5cm}
     \begin{subfigure}[b]{0.43\textwidth}
         \centering
         \begin{tikzpicture}
         \draw (0, 0) node[inner sep=0] {\includegraphics[width=\textwidth,trim={15mm 0 15mm 0},clip]{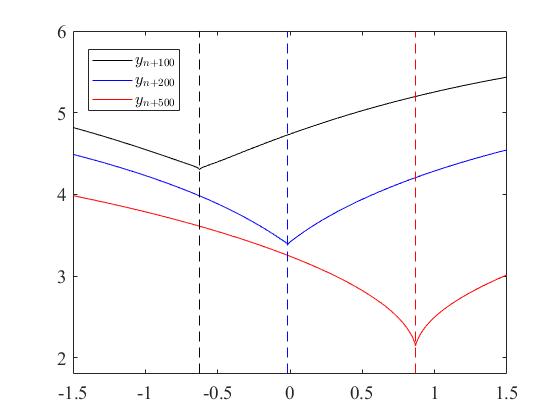}};
         \draw (-2.5, -2) node {(B)};
         \end{tikzpicture}
         \label{fig:y equals x}
     \end{subfigure}
        \caption{\raggedright Entanglement entropy vs. predicted function values ($y_{n+k}$ for the $k$-th iteration) within the search range for sawtooth functions with amplitude 1. The figures depict the results of the 100th, 200th, and 500th iterations. (A) Noiseless sawtooth function. (B) Sawtooth function with normally distributed noise.}
        \label{app1}
\end{figure}
This phenomenon is exemplified in the illustration provided in Fig.~\ref{app1}. In this instance, we apply our new method to extrapolate the sawtooth function for 512 points, both with and without noise. The figures showcase how the entanglement entropy varies when our algorithm employs different estimations, aside from the optimal one, at the 100th, 200th, and 500th iterations. Notably, the entropy experiences increasingly accelerated decay as it approaches the minimum, with the sharp tip at the minimum indicating a precise point for the estimated result.
\begin{figure}[htb!]
     \centering
     \begin{subfigure}[b]{0.43\textwidth}
         \centering
         \begin{tikzpicture}
         \draw (0, 0) node[inner sep=0] {\includegraphics[width=\textwidth,trim={15mm 0 15mm 0},clip]{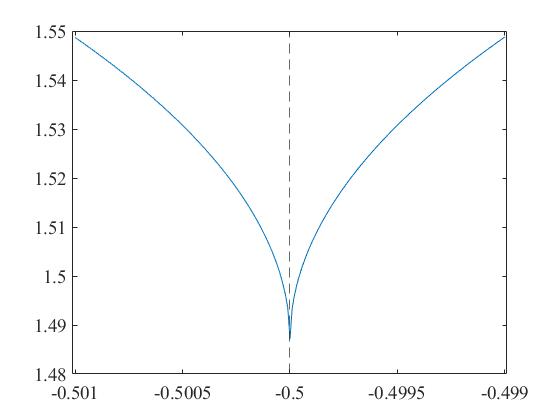}};
         \draw (-2.5, -1.5) node {(A)};
         \end{tikzpicture}
         \label{fig:y equals x}
     \end{subfigure}
     \hspace{0.5cm}
     \begin{subfigure}[b]{0.43\textwidth}
         \centering
         \begin{tikzpicture}
         \draw (0, 0) node[inner sep=0] {\includegraphics[width=\textwidth,trim={15mm 0 15mm 0},clip]{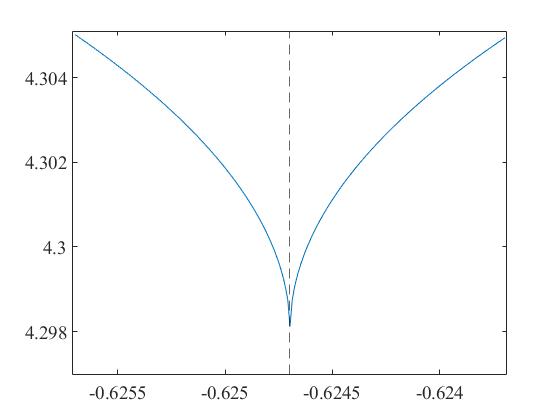}};
         \draw (-2.5, -1.5) node {(B)};
         \end{tikzpicture}
         \label{fig:y equals x}
     \end{subfigure}
        \caption{\raggedright Closer view around minimum entanglement entropy of $y_{n+100}$ results in Fig.~\ref{app1}. (A) Noiseless. (B) With normally distributed noise.}
        \label{app2}
\end{figure}
In Fig.~\ref{app2}, we present a higher precision visualization of the entropy. The pronounced tip at the minimum entropy might initially seem counterintuitive, as values in proximity to the optimal point should also possess relatively high credibility. This observation corresponds to a non-convex feature around the minimum, which is an intriguing aspect of the entropy behavior.

Analyzing entropy behaviors serves to authenticate the efficacy of our pursuit of minimal entropy. The paramount aim of our algorithm is to pinpoint the absolute minimum within the search range, rather than a mere local minimum. The algorithm's success hinges upon its capacity to precisely locate this absolute minimum; any failure to do so renders the algorithm ineffective. The entropy behaviors allow us to affirm the algorithm's triumph in the given example, as the predicted optimal point aligns perfectly with the local minimum.

We conducted a comprehensive examination of entropy behaviors across a diverse collection of functions, encompassing plane waves, square waves, and more. Across most instances of entropy extrapolation, the entropy exhibits a similar pattern to that of the sawtooth function, albeit with minor distinctions attributable to the function type and the presence of noise. For instance, as illustrated in Fig.~\ref{app1}, the addition of noise to the input data results in a broadening of the entropy curves. We also encountered cases where noise imparts a slight asymmetry to the curve in relation to the optimal $y$ value corresponding to minimal entropy. Despite these slight variations, the estimation outcomes generated by entropy extrapolation remain valid. This is attributed to the fact that these factors do not exert significant alterations on the entanglement entropy manifested by the functions.

\begin{figure}[htb!]
     \centering
     \begin{subfigure}[b]{0.43\textwidth}
         \centering
         \begin{tikzpicture}
         \draw (0, 0) node[inner sep=0] {\includegraphics[width=\textwidth,trim={15mm 0 15mm 0},clip]{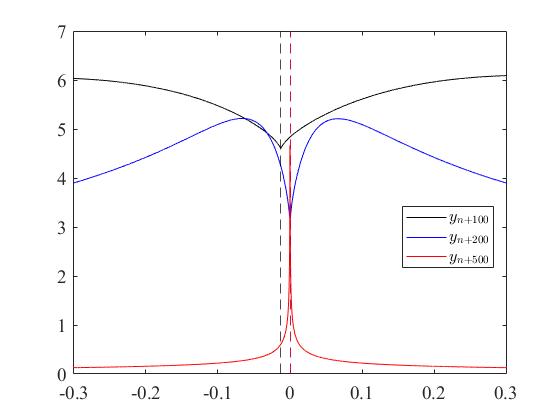}};
         \draw (-2.5, -1.5) node {(A)};
         \end{tikzpicture}
         \label{fig:y equals x}
     \end{subfigure}
     \hspace{0.5cm}
     \begin{subfigure}[b]{0.43\textwidth}
         \centering
         \begin{tikzpicture}
         \draw (0, 0) node[inner sep=0] {\includegraphics[width=\textwidth,trim={15mm 0 15mm 0},clip]{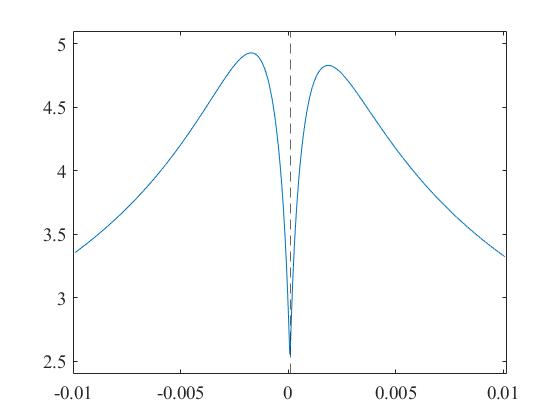}};
         \draw (-2.5, -1.5) node {(B)};
         \end{tikzpicture}
         \label{fig:y equals x}
     \end{subfigure}
        \caption{\raggedright Entanglement entropy vs. $y_{n+k}$ for the half circle in the frequency domain (the example in Fig.~\ref{fig:2}A). (A) Global view across the search range. Results of the 100th, 200th, and 500th iterations are provided. (B) The specific view around the optimal point given by the algorithm at the 500th iteration.}
        \label{app4}
\end{figure}

However, in exceptionally rare instances, the entropy exhibits unconventional behavior around the optimal point indicated by entropy extrapolation. In such cases, the validity of estimation results for future function values ceases to hold. Fig.~\ref{app4} illustrates the entropy curves at various iterations during the extrapolation of the half-unit circle example $f(x) = \mathcal{F}^{-1}[\sqrt{1-(\omega-3)^2}]$. As depicted in the figure, the entropy no longer converges towards an absolute minimum within the search range, indicating the absence of a preferred estimation that the algorithm can produce. These particular functions fall outside the realm of well-performing scenarios and cannot be effectively predicted by our algorithm. To circumvent the application of entropy extrapolation to functions residing in unfavorable regions, a critical step involves scrutinizing the entropy behavior post the application of entropy extrapolation.

\label{appdx}

\newpage
\nocite{*}
\bibliography{Ref}

\end{document}